\documentclass[12pt,preprint]{aastex}          % preprint format, one column
   \usepackage{graphicx}
   \usepackage{amssymb}
   \usepackage{epstopdf}

\def\teq#1{$\, #1\,$}
{\catcode`\@=11                                                  
   \gdef\SchlangeUnter#1#2{\lower2pt\vbox{\baselineskip 0pt\lineskip0pt    
   \ialign{$\m@th#1\hfil##\hfil$\crcr#2\crcr\sim\crcr}}}}           
\def\gtrsim{\mathrel{\mathpalette\SchlangeUnter>}}               
\def\lesssim{\mathrel{\mathpalette\SchlangeUnter<}}
          \font\sixrm=cmr6       
                     
\def\pr{Phys. Rev.}                             % DO NOT DELETE
\def\apss{Astr. Space Sci.}                     % DO NOT DELETE
\def\asr{Adv. Space Res.}                       % DO NOT DELETE
\def\prl{Phys. Rev. Lett.}                      % DO NOT DELETE
\def\ssr{Space Sci. Rev.}                       % DO NOT DELETE

\def\ThetaBnone{\Theta_{\hbox{\sixrm Bn1}}} 
\def\ThetaBntwo{\Theta_{\hbox{\sixrm Bn2}}} 
\def\Machson{{\cal M}_{\hbox{\sixrm S}}}   
\def\Machalf{{\cal M}_{\hbox{\sixrm A}}}   
\def\Chandra{{\it Chandra}}  
\begin{document}
\newcommand{\vol}[2]{$\,$\rm #1\rm , #2.}                 
\newcommand{\figureoutpdf}[5]{\centerline{}
   \centerline{\hspace{#3in} \includegraphics[width=#2truein]{#1}}
   \vspace{#4truein} \caption{#5} \centerline{} }
\newcommand{\twofigureoutpdf}[3]{\centerline{}
   \centerline{\includegraphics[width=2.9truein]{#1}
        \hspace{0.4truein} \includegraphics[width=2.9truein]{#2}}
        \vspace{-0.1truein}
        \caption{{\small #3}} \vspace{-0.15truein}}    % plot side-by-side PDF version for TeXShop

\title{Electrostatic Potentials in Supernova Remnant Shocks}
%\title{ELECTROSTATIC POTENTIALS IN SUPERNOVA REMNANT SHOCKS}

   \author{Matthew G. Baring and Errol J. Summerlin}
   \affil{Department of Physics and Astronomy MS-108, \\
      Rice University, P.O. Box 1892, Houston, TX 77251, U.S.A.\\
      \tt baring@rice.edu, xerex@rice.edu\rm}

\slugcomment{Accepted for publication in
             \it Astrophysics and Space Science\rm}

\begin{abstract}  
Recent advances in the understanding of the properties of supernova
remnant shocks have been precipitated by the \Chandra\ and XMM X-ray
Observatories, and the HESS Atmospheric \v{C}erenkov Telescope in the
TeV band.  A critical problem for this field is the understanding of the
relative degree of dissipative heating/energization of electrons and
ions in the shock layer.  This impacts the interpretation of X-ray
observations, and moreover influences the efficiency of injection into
the acceleration process, which in turn feeds back into the thermal
shock layer energetics and dynamics.  This paper outlines the first
stages of our exploration of the role of charge separation potentials in
non-relativistic electron-ion shocks where the inertial gyro-scales are
widely disparate, using results from a Monte Carlo simulation.  Charge
density spatial profiles were obtained in the linear regime, sampling
the inertial scales for both ions and electrons, for different magnetic
field obliquities.  These were readily integrated to acquire electric
field profiles in the absence of self-consistent, spatial readjustments
between the electrons and the ions.  It was found that while diffusion
plays little role in modulating the linear field structure in highly
oblique and perpendicular shocks, in quasi-parallel shocks, where charge
separations induced by gyrations are small, and shock-layer electric
fields are predominantly generated on diffusive scales.
\end{abstract}

\section{Introduction}
 \label{sec:intro}

The understanding of the character of shells and interiors of supernova
remnants (SNRs) has been advanced considerably by groundbreaking
observations by the \Chandra\ X-ray Observatory.  These have been
enabled by its spectral resolution coupled with its impressive angular
resolution.  Of particular interest to the shock acceleration and cosmic
ray physics communities is the observation of extremely narrow
non-thermal X-ray spatial profiles in selected remnants (see Long et al.
2003 for SN1006; Vink \& Laming 2003 for Cas A; and Ellison \&
Cassam-Chena\"i 2005, for theoretical modeling), which define strong
brightness contrasts between the shell, and the outer, upstream zones.
If the synchrotron mechanism is responsible for this non-thermal
emission, the flux ratios from shock to upstream indicate strong
magnetic field enhancement near the shock.  These ratios considerably
exceed values expected for hydrodynamic compression at the shocked
shell, so proposals of magnetic field amplification (e.g. Lucek \& Bell
2000) in the upstream shock precursor have emerged.

Another striking determination by \Chandra\ concerns electron heating by
ions in the shock layer.  Dynamical inferences of proton temperatures in
remnant shocks can be made using proper motion studies of changes in a
remnant's angular size, or more direct spectroscopic methods (e.g.
Ghavamian et al. 2003). In the case of remnant 1E 0101.2-7129, Hughes et
al. (2000) used a combination of ROSAT and Chandra data spanning a
decade to deduce an expansion speed. Electron temperatures \teq{T_e} are
determined using ion line diagnostics (assuming the equilibration
\teq{T_e=T_p}), via both the widths and relative strengths for different
ionized species.  From these two ingredients, Hughes et al. (2000)
observed that deduced proton temperatures were considerably cooler, i.e.
\teq{3kT_p/2 \ll m_p (3u_{1x}/4)^2 /2}, than would correspond to
standard heating for a strong hydrodynamic shock with an upstream flow
speed of \teq{u_{1x}}.  The same inference was made by Decourchelle et
al. (2000) for Kepler's remnant, and by Hwang et al. (2002) for Tycho's
SNR.  This property is naturally expected in the nonlinear shock
acceleration scenario that is widely used in describing cosmic ray and
relativistic electron generation in SNRs:  the highest energy particles
tap significant fractions of the total available energy, leading to a
reduction in the gas temperatures. This nonlinear hydrodynamic
modification has been widely discussed in the cosmic ray acceleration
literature (e.g. see Jones \& Ellison 1991; Berezhko \& Ellison 1999,
and references therein), and has been extensively applied to
multiwavelength SNR spectral models (e.g. see Baring et al. 1999;
Berezhko et al. 2002; Ellison \& Cassam-Chena\"i 2005; Baring, Ellison
\& Slane 2005).

The extent of equilibration between electrons and ions in SNR shell
shocks needs to be understood, and can potentially be investigated by
laboratory plasma experiments.  A critical ingredient is the
electrostatic coupling between electrons and protons in the shock layer,
which offers the potential for considerable heating of \teq{e^-},
coupled with cooling of protons, setting \teq{m_p (3u_{1x}/4)^2 /2 \gg
3kT_e/2\gg m_e (3u_{1x}/4)^2 /2} with \teq{T_e\neq T_p}.  Probing this
coupling is the subject of this paper.  Here we describe preliminary
results from our program to explore electrostatic energy exchange
between these two species in SNR shocks, using a Monte Carlo simulation
of charged particle transport, their spatial distribution and associated
electric field generation. The goal is to eventually obtain a simulation
with self-consistent feedback between the charge separation potentials,
and the Lorentz force they impose on the charges. The research progress
outlined here indicates that the role of diffusion in quasi-parallel
shocks is very important, and can readily influence computed charge
separation potentials.

\section{Shock Layer Electrostatics in Supernova Remnants}

Cross-shock electrostatic potentials arise in the shock layer because of
the different masses of electrons and ions: upstream thermal ions gyrate
on larger scales than do their electron counterparts when they transit
downstream of the shock for the first time.  In shocks where the field
is oblique to the shock normal by some angle \teq{\ThetaBnone} upstream
(and therefore a greater angle downstream), on average, protons will be
located further downstream of the shock than electrons.  This naturally
establishes an electric field {\bf E}, to which the plasma responds by
accelerating electrons and slowing down ions to short out the induced
{\bf E}.  A feedback loop ensues, mediated by fields and currents that
vary spatially on the order of, or less than, the ion inertial scale,
which is typically shorter than the ion gyroradius in astrophysical
shocks such as those associated with supernova remnant shells.

Particle-in-cell (PIC) simulations are a natural technique (e.g. see
Forslund \& Friedberg 1971 for an early implementation) for exploring
signatures of such electrostatics in shock layers.  These trace particle
motion and field fluctuations, obtained as self-consistent solutions of
the Newton-Lorentz and Maxwell's equations, in structured zones or cells
in spatially-constrained boxes.  Such simulations have been used
recently to probe the Weibel instability at weakly-magnetized,
perpendicular, relativistic pair plasma shocks (see Silva et al. 2003;
Hededal et al. 2004; Nishikawa et al. 2005).  They have also been used
by Shimada \& Hoshino (2000) to treat electrostatic instabilities at
non-relativistic, quasi-parallel electron-ion shocks.  Rich in their
turbulence information, due to their intensive CPU requirements, such
simulations have difficulty in modeling realistic \teq{m_p/m_e} mass
ratios, and fully exploring 3D shock physics such as diffusive
transport.  Moreover, they cannot presently address the wide range of
particle momenta and spatial/temporal scales encountered in the
acceleration process; they often do not obtain time-asymptotic states
for the particle distributions.

Monte Carlo techniques provide an alternative method that can easily
resolve electron and proton inertial scales, treat fully 3D transport
and large dynamical ranges in spatial and momentum scales, all at modest
computation cost.  While they parameterize the effects of turbulence via
diffusive mean free paths (e.g. see Jones \& Ellison 1991), they can
accurately describe the microphysics of cross-shock electrostatic
potentials. This simulational approach has been well-documented in the
literature (e.g. Jones \& Ellison 1991; Ellison, Jones \& Baring 1996),
with definitive contributions to the study of heliospheric shock
systems, cosmic ray production, SNR applications and gamma-ray bursts.
It models the convection and diffusion of gyrating particles in
spatially-structured flows and fields, with transport back and forth
across the shock effecting diffusive Fermi-type acceleration directly
from the thermal population. The mean free path \teq{\lambda} is usually
prescribed as some increasing function of particle momentum \teq{p} or
gyroradius \teq{r_g}.  Here we use this approach, with
\teq{\lambda\propto p} adopted as a broadly representative situation:
see Baring et al. (1997) for a discussion of evidence from observations 
and plasma simulations in support of such a specialization. Here,
\teq{\lambda /r_g=5} is chosen for illustrative purposes, to begin to
investigate electrostatic influences on thermal and low-energy
non-thermal particles in non-relativistic electron-ion shocks of
arbitrary \teq{\ThetaBnone}.  Clearly then, the diffusive scales
for protons and electrons are disparate by their mass ratio.

In the Monte Carlo simulation, the shock is defined
magnetohydrodynamically, consisting of laminar, uniform flows and fields
upstream and downstream of a sharp, planar discontinuity.  The magnetic
fields and flow velocities either side of the shock are related uniquely
through the standard Rankine-Hugoniot solutions for energy and momentum
flux conservation (e.g. see Boyd \& Sanderson 1969). These solutions
include essential elements of Maxwell's equations, such as the
divergenceless nature of the {\bf B} field.  In the reference frames of
the local upstream and downstream fluids, the mean electric field is
assumed to be zero, a consequence of very effective Debye screening, so
that the only electric fields present in shock rest frames are {\bf
u\teq{\times}B} drift fields. The charged electrons and protons (more
massive ions are omitted in this paper to simplify the identification of
the principal effects) are treated as test particles, convecting into
and through the shock, initially with the prescribed upstream fluid
velocity {\bf u}$_1$.  This neutral beam is entirely thermal, and moreover is
in equipartition, so that it has an input temperature  \teq{T_e=T_p}.  
The charges constantly diffuse in space to mimic collisions with
magnetic turbulence that is putatively present in the shock environs, and
in so doing, can be accelerated.  These non-thermal particles form a minority
of the total population, and provide only a minor contribution to the 
fields illustrated in this paper.

The charges transiting the shock  distribute their downstream density in
a manner that couples directly to their gyrational motion (e.g. see Baring 2006),
and the local densities of electrons and protons can easily be tracked
in the Monte Carlo technique by accumulating ``detection'' data at
various distances from the shock.  Monte Carlo simulation runs clearly
exhibit non-zero charge excursions within a proton gyroradius of the
shock, an effect similar to those found in PIC codes. For example, a
cold, neutral \teq{e-p} upstream beam develops an electron concentration
near the shock in the downstream region, with protons distributed on
their larger inertial scales.  The resulting charge distributions
\teq{\rho (x)} depend on both the upstream field obliquity
\teq{\ThetaBnone}, and also on the sonic Mach number \teq{\Machson
\approx u_{1x}/\sqrt{5kT_p/(3m_p)}} in situations where the upstream
beam is warm.  Due to the steady-state, planar nature of the simulation,
these distributions depend only on the coordinate $x$ along the shock
normal. It is straightforward using Gauss' law for electrostatics,
\teq{\mathbf{\nabla \cdot E} =4\pi\rho (x)}, to integrate the charge
distribution profile to obtain \teq{E_x(x)= -\partial\Phi /\partial x}.
Eventually, such ``{\it linear}'' fields will then be used to compute
the energy exchange between electrons and ions as they cross the
non-monotonic charge separation potential \teq{\Phi (x)}.

\begin{figure*}[t]
\twofigureoutpdf{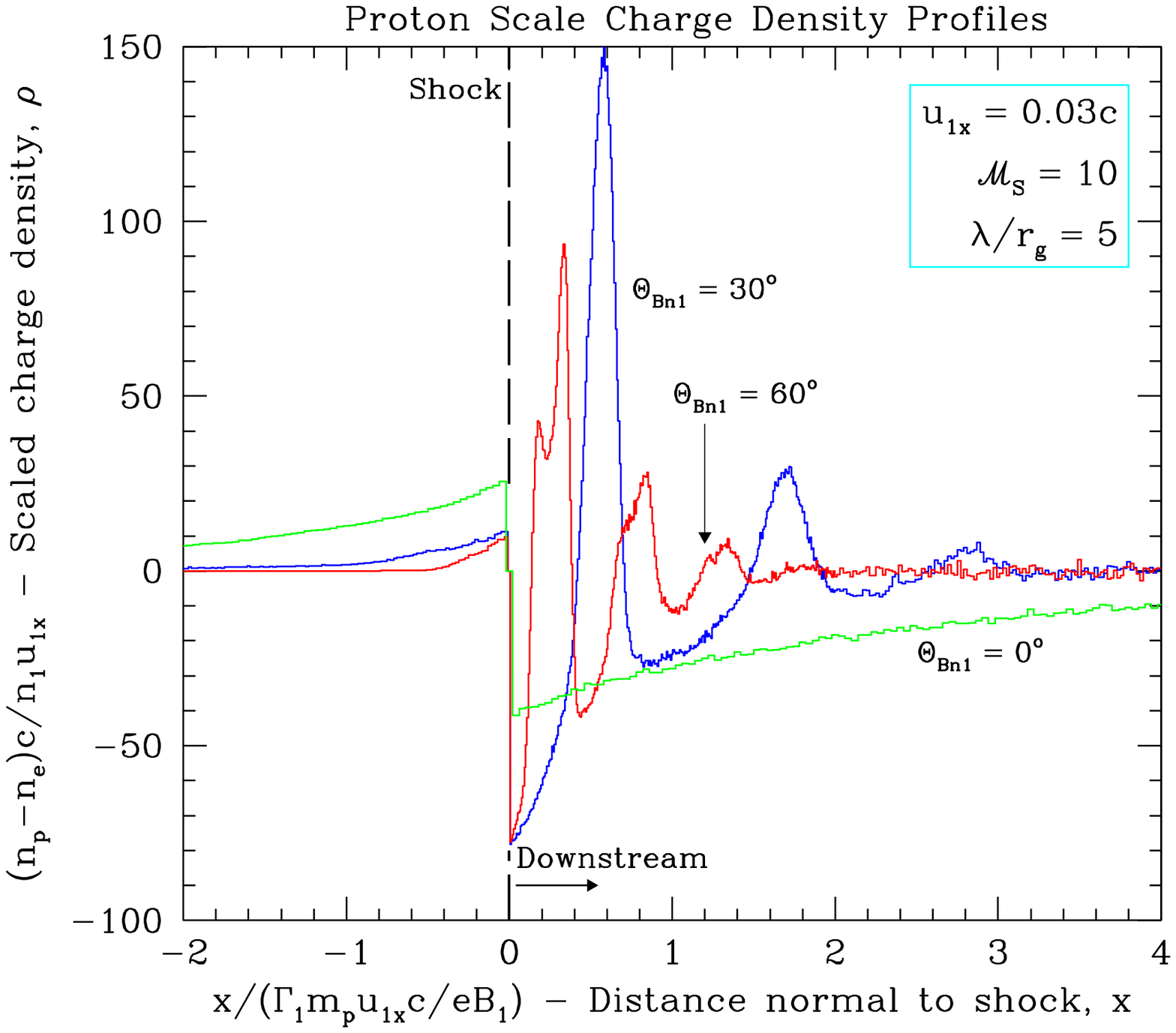}{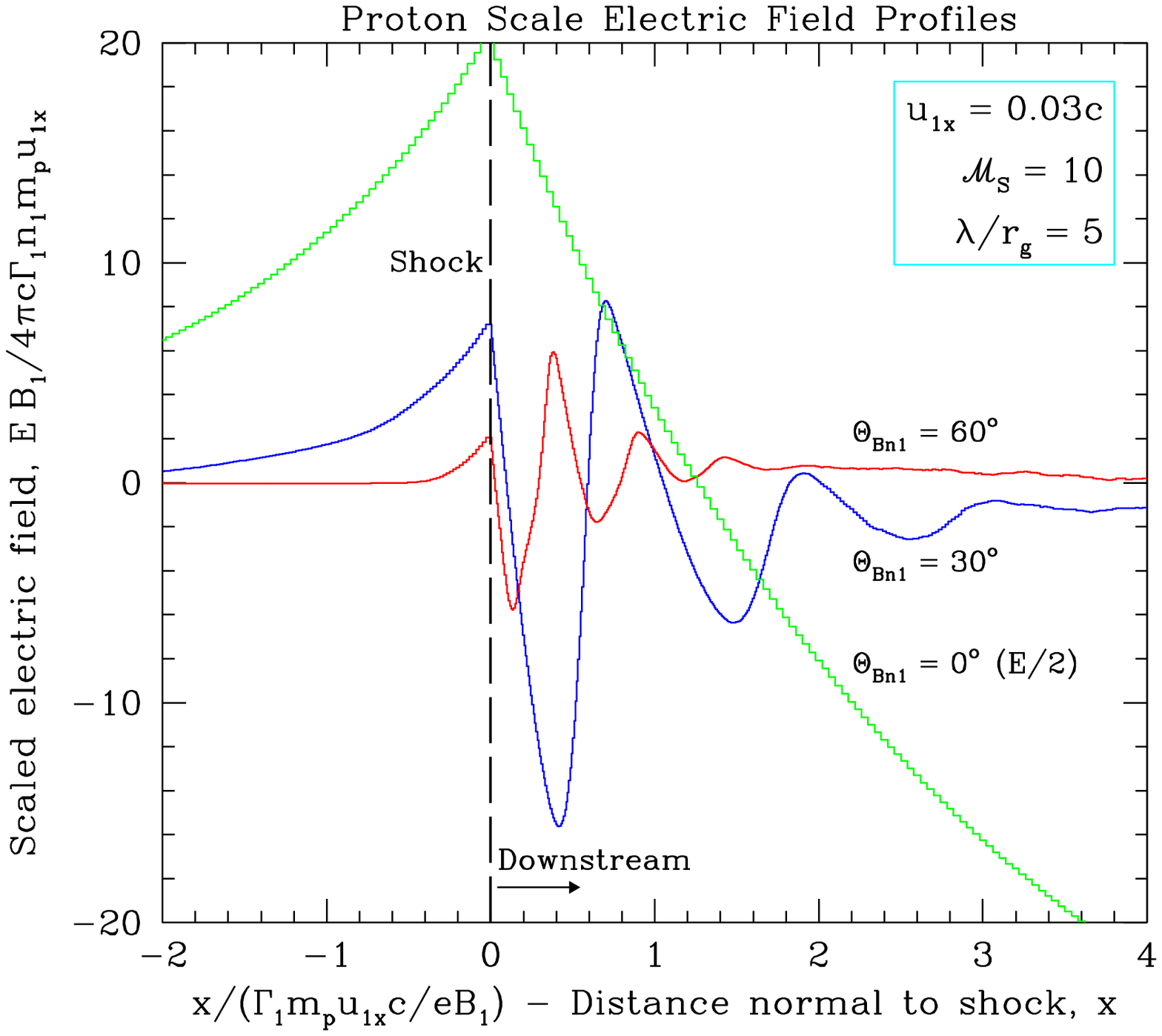}{
Electrostatic profiles for an electron-proton plasma shock of speed
\teq{u_{1x}=9000}km/sec and sonic Mach number \teq{\Machson =10}, with
upstream field obliquities \teq{\ThetaBnone=0^{\circ}} (green),
\teq{\ThetaBnone=30^{\circ}} (blue) and \teq{\ThetaBnone=60^{\circ}}
(red), as labelled.   {\it Left panel}: scaled charge density
distribution \teq{\rho (x)}; {\it Right Panel}: resulting ``{\it
linear}'' electric field profile \teq{E\equiv E_x(x)} computed by
solving Gauss' Law (the \teq{\ThetaBnone=0^{\circ}} displays
\teq{E_x(x)/2}). The profiles are exhibited on proton gyroscales
\teq{r_{g,p}=m_pu_{1x}c/eB_1}, so that fluctuations on electron
gyroscales are collapsed into the shock layer (dashed vertical line).   
The panels display both oscillations associated with proton gyrations in
oblique cases, and diffusive upstream ``precursors,'' which are most
prominent when \teq{\ThetaBnone=0^{\circ}} (a parallel shock).
 \label{fig:electrostatic} }      %fig. 1
\end{figure*}

Linear determinations of electrostatic spatial profiles are shown in
Figure~\ref{fig:electrostatic} to illustrate the key features; these did
not self-consistently include the acceleration of electrons and protons
in the produced {\bf E} field.  The left panel depicts large charge
density fluctuations that trace ion gyration in the downstream magnetic
field. Similar fluctuations of opposite sign are created by electrons,
but on much smaller scales that are not resolved in the Figure. 
Accordingly, ``striped'' zones of positive and negative charge density
result, and this electrostatic analog of a plasma oscillation integrates
to produce the {\bf E} fields in the right panel that can accelerate or
decelerate electrons and protons.  The outcome depends on the shock
obliquity \teq{\ThetaBnone} when \teq{\ThetaBnone\lesssim 60^{\circ}},
whereas quasi-perpendicular shocks with \teq{\ThetaBnone\gtrsim
60^{\circ}} possess profiles fairly close to the \teq{\ThetaBnone
=60^{\circ}} case depicted in Figure~\ref{fig:electrostatic}, since they
all have field obliquities \teq{\ThetaBntwo\approx 80^{\circ}} -
\teq{90^{\circ}} downstream. Note also, that while the gyrational
contributions are prominent, there is also a diffusive contribution,
manifested as an upstream {\it precursor} modification to \teq{\rho (x)}
and {\bf E}.  This is particularly marked in the parallel shock
(\teq{\ThetaBnone =0^{\circ}}) case, where the diffusive scale along the
field achieves a maximal component orthogonal to the shock plane. This
diffusive influence originates in accelerated particles returning to the
upstream side of the shock (\teq{x<0}), enhancing the density there
before convecting downstream again: protons effect this on larger
scales, and so control the precursors seen in the Figure (i.e. \teq{\rho
(x)>0} for \teq{x < 0}).  Since the fields are established on the scale
of a proton gyroradius, their magnitude scales as \teq{E_x\sim 4\pi\rho
r_{g,p} =4\pi e n_p (m_pu_{1x}c/eB_1)}, so that \teq{\vert E_x/B_1\vert
\sim 4\pi n_pm_pu_{1x}c/B_1^2\equiv \Machalf^2 (c/u_{1x})\gg 1} for
Alfv\'enic Mach numbers \teq{\Machalf > 1}.

The competition between gyrational and diffusive influences on
electrostatics is a principal conclusion of this paper, defining a
dichotomy delineating quasi-parallel and quasi-perpendicular shocks. The
Monte Carlo technique can accurately trace both influences, while
comfortably resolving the disparate scales for the $e$-$p$ shock
problem.   Since the ``linear'' results illustrated need to be upgraded
to account for the {\bf E}-field's influence on \teq{e^-} and \teq{p}
motions, it is presently unclear whether ions can energize electrons
overall (the right panel of the Figure suggests they may even decelerate
them), and how the net work done depends on field obliquity.  A
noticeable feature of the electric field profiles in
Fig.~\ref{fig:electrostatic} is that for \teq{\ThetaBnone \lesssim
60^{\circ}}, these linear field calculations do not establish \teq{\vert
\hbox{\bf E} \vert \to 0} asymptotically as \teq{\vert x\vert\to\infty},
as required by net charge neutrality. The next step of this program will
be to solve the Newton-Lorentz equation of motion \teq{d\hbox{\bf p}/dt
= q(\hbox{\bf E} + \hbox{\bf v}\times \hbox{\bf B}/c)} to determine both
drift and accelerative contributions to the charges' motions.   These
will necessitate a recomputation of the {\bf E} field profiles, and a
feedback loop will result, with shock layer currents generating magnetic
field excursions via Amp\`ere's law, \teq{\mathbf{\nabla \times B} =4\pi
\mathbf{J}/c}. This iterative process will continue to convergence
(establishing \teq{\vert \hbox{\bf E} \vert \to 0} as \teq{\vert
x\vert\to\infty}), with relaxation to equilibrium occuring on the
spatial response scale \teq{u_{1x}/\omega_p}, where \teq{\omega_p =
\sqrt{4\pi e^2n_p/m_p}} is the proton plasma frequency.  Since
\teq{u_{1x}/(\omega_p r_{g,p})\sim u_{1x}/(c\Machalf ) \ll 1}), this
response scale is far less than a proton gyroradius for typical SNR 
environmental parameters, and indeed for any strong, non-relativistic 
astrophysical shock.

The degree of electron energization in the cross shock potential may
offer significant insights into the well-known electron injection
problem at non-relativistic shocks. Electrons do not resonantly interact
with Alfv\'en waves until they become relativistic. Levinson (1992) suggested
that \teq{e^-} interaction with a presumably abundant supply of whistler
waves could effect pre-injection into diffusive acceleration processes,
if electrons could achieve energies in excess of around 10 keV to access
the whistler resonance branch.  The planned self-consistent extension of
the developments outlined here will help determine whether this channel
of access to continued acceleration is opened up by shock layer
electrostatics.  Moreover, crafted laboratory plasma experiments might
cast light on this aspect of shock layer physics.

\newpage

\section{Conclusion}

In this paper, charge density and associated cross-shock electric field
spatial profiles are presented for different magnetic field obliquities.
It was found that in highly oblique and perpendicular shocks diffusion
plays little role in modulating the field structure, which is controlled
by the magnetic kinking and compression on the downstream side of the
shock.  In contrast, in quasi-parallel shocks, where the gyrational
charge separation is small, diffusion scales upstream and downstream of
the shock dominate the generation of shock-layer electric fields.  This
is an interesting twist, suggesting that observationally, thermal X-ray
emission could be distinctly different in portions of an SNR rim that
establish quasi-parallel and quasi-perpendicular shocks.  The work
discussed here paves the way for self-consistent determination of the
acceleration/deceleration of electrons and protons, their spatial
distributions, and the electric fields normal to non-relativistic
shocks.  This development will impact the understanding of electron
injection and acceleration in shocks of all obliquities.

\vskip 6pt
\noindent {\bf References:}
\vskip 2pt
\def\ref{\noindent \hangafter=1 \hangindent=0.7 true cm}

\def\mn{M.N.R.A.S.}
\def\aas{{Astron. Astrophys.}}
\def\aassupp{{Astron. Astrophys. Supp.}}
\def\apss{{Astr. Space Sci.}}
\def\apj{ApJ}
\def\nat{Nature}
\def\aaps{{Astron. \& Astr. Supp.}}
\def\xviiicrc{{17th Internat. Cosmic Ray Conf. Papers}}
\def\xviiiicrc{{18th Internat. Cosmic Ray Conf. Papers}}
\def\xixicrc{{19th Internat. Cosmic Ray Conf. Papers}}
\def\xxicrc{{20th Internat. Cosmic Ray Conf. Papers}}
\def\xxiicrc{{21st Internat. Cosmic Ray Conf. Papers}}
\def\apj{{ApJ}}
\def\aa{{A\&A}}
\def\apjs{{ApJS}}
\def\sp{{Solar Phys.}}
\def\jgr{{J. Geophys. Res.}}
\def\grl{{Geophys. Res. Lett.}}
\def\jphysb{{J. Phys. B}}
\def\ssr{{Space Science Rev.}}
\def\araa{{Ann. Rev. Astron. Astrophys.}}
\def\nature{{Nature}}
\def\asr{{Adv. Space. Res.}}
\def\prc{{Phys. Rev. C}}
\def\prd{{Phys. Rev. D}}
\def\pr{{Phys. Rev.}}
\def\rgph{{Rev. Geophys.}}
\def\cjp{{Canadian J. Physics}}
\def\np{{Nuclear Physics}}
\def\sturrock{{Solar Flares}, ed. P. A. Sturrock,
       (Boulder:Colorado Associated University Press)}
\def\accaip{{Particle Acceleration Mechanisms in 
Astrophysics}, ed. J.~Arons, C.~Max, C.~McKee (New York: AIP)}
\def\ljaip{{Gamma Ray Transients and Related Astrophysical Phenomena,
            } eds. R. E. Lingenfelter, H. S. Hudson, and D. M. Worrall
            (New York:AIP)}
\def\gsfcaip{{Positron Electron Pairs in Astrophysics},
    eds. M. L. Burns, A. K. Harding, and R. Ramaty (New York:AIP)}
\def\washaip{{Nuclear Spectroscopy of Astrophysical
    Sources}, eds. N. Gehrels and G. H. Share, (New York:AIP)}
\def\minaip{{Cosmic Abundances of Matter}, ed. C. J. Waddington,
(New York:AIP)}
\def\gcacta{{Geochim. Cosmochim. Acta}}
\def\mnras{{M.N.R.A.S.}}
\def\ref{\noindent\hangafter=1\hangindent=1truecm}

\small
{\baselineskip 11pt  \parskip 4pt

\ref
Baring, M.~G. 2006, on-line proceedings of the 2006 KITP/UCSB
conference ``Supernova and Gamma-Ray Burst Remnants'' 
[{\tt http://online.kitp.ucsb.edu/online/grb\_c06/baring/}]

\ref
Baring, M.~G., Ellison, D.~C., Reynolds, S.~P., Grenier, I.~A., \& Goret, P.
1999, \apj,\vol{513}{311}

\ref
Baring, M.~G., Ellison, D.~C., \& Slane, P.~O. 2005,  \asr,\vol{35}{1041}

\ref
Baring, M. G., Ogilvie, K. W., Ellison, D. C., \& Forsyth, R. J. 1997,
	\apj,\vol{476}{889}

\ref
Berezhko, E.~G. \& Ellison, D.~C. 1999, \apj,\vol{526}{385}

\ref
Berezhko, E.~G., Ksenofontov, L.~T. \& V\"olk, H.~J. 2002, \apj,\vol{395}{943}

\ref
Boyd, T.~J.~M. \& Sanderson, J.~J. 1969, {\it Plasma Dynamics}, 
(Nelson \& Sons, London)
   
\ref
Decourchelle, A., Ellison, D.~C. \& Ballet, J. 2000, \apj,\vol{543}{L57} % Kepler

\ref
Ellison, D.~C., Baring, M.~G. \& Jones, F.~C. 1996, \apj,\vol{473}{1029}

\ref
Ellison, D.~C. \& Cassam-Chena\"i, G. 2005, \apj,\vol{632}{920}

%\ref
%Fesen, R.~A., et al. 2006, \apj,\vol{636}{859}

\ref
Forslund, D.~W. \& Freidberg, J.~P. 1971, \prl\vol{27}{1189}

\ref
Ghavamian, P., Rakowski, C.~E., Hughes, J.~P. \& Williams, T.~B. 2003, \apj,\vol{590}{833}

\ref
Hededal,ÊC.~B., Haugbolle,ÊT., Frederiksen,ÊJ.~T. \& Nordlund,ÊA 2004,
\apj,\vol{617}{L107}

\ref
Hughes, J.~P., Rakowski, C.~E., \& Decourchelle, A. 2000, \apj,\vol{543}{L61} % 1E0102

\ref
Hwang, U., et. al. 2002, \apj,\vol{581}{110}  % Tycho

%\ref
%Hwang, U., et. al. 2004, \apj,\vol{615}{L117} % Cas A

\ref
Jones, F.~C. \& Ellison, D.~C. 1991, \ssr,\vol{58}{259}

\ref
Levinson, A. 1992, \apj,\vol{401}{73}

\ref
Long, K.~S., et al. 2003, \apj,\vol{586}{1162}

\ref
Lucek, S.~G. \&  Bell, A.~R. 2000, \mnras,\vol{314}{65}

\ref
Nishikawa, K.-I., et al. 2005, \apj,\vol{622}{927}

\ref
Shimada, N. \& Hoshino, M. 2000, \apj,\vol{543}{L67}
   
\ref
Silva, L.~O., et al. 2003, \apj,\vol{596}{L121}

\ref
Vink, J. \& Laming, J.~M. 2003, \apj,\vol{584}{758} % Cas A

}

\end{document}